\documentclass[12pt]{article}
\usepackage[margin=1in]{geometry}
\usepackage{amsmath,amssymb,bm,mathtools}
\usepackage{booktabs,graphicx,float,longtable,array}
\usepackage[dvipsnames]{xcolor}
\usepackage{microtype}
\usepackage[hidelinks]{hyperref}
\usepackage{enumitem,setspace,xspace,caption}
\usepackage[round]{natbib}
\usepackage[affil-it]{authblk}
\usepackage{kotex}
\usepackage{ulem}

\graphicspath{{figures/}}

\title{Bayesian Generalized Network Autoregressive Model with
Structured Shrinkage and Persistence Priors}
\author[1]{Seongmin Kim}
\author[2]{Kyusoon Kim$^{*}$}

\affil[1]{Human-Centered Artificial Intelligence Research Institute,
Ewha Womans University, Seoul 03760, Korea}
\affil[2]{Department of Statistics and Actuarial Science,
Soongsil University, Seoul 06978, Korea}
\date{August 2026}

\begin{document}
\maketitle
\begingroup
\renewcommand\thefootnote{\fnsymbol{footnote}}
\footnotetext[1]{Corresponding author: kyusoon.kim@ssu.ac.kr}
\endgroup

\begin{abstract}
    We propose a Bayesian generalized network autoregressive (BGNAR) model for multivariate time series observed over a network. The model combines the parsimonious structure of the generalized network autoregressive (GNAR) model with shrinkage and persistence priors adapted from Bayesian vector autoregressive (BVAR) modeling. Minnesota-type priors shrink the own-lag and network-lag coefficients more strongly at higher temporal lags, with additional shrinkage across neighborhood orders for network effects. A hierarchical prior allows node-specific own-lag coefficients to share information across nodes while retaining heterogeneity. We also adapt the sum-of-coefficients and dummy-initial-observation priors to the GNAR parameterization. Simulation studies show that BGNAR can accommodate a fixed over-specified temporal and neighborhood structure without dataset-specific BIC order selection, while achieving forecasting accuracy comparable to GNAR and outperforming BVAR in the settings considered. The Bayesian formulation additionally provides posterior inference for dynamic effects and predictive uncertainty. We further illustrate the proposed method using a wind-speed network application.
\end{abstract}

\noindent\textbf{Keywords:} Network time series; Generalized network autoregressive model; Bayesian vector
autoregression; Hierarchical shrinkage; Posterior predictive inference.

\section{Introduction}
Multivariate time series are increasingly observed over collections of interacting units whose relationships can be represented by a network. Such data are commonly referred to as network time series. Examples include measurements collected at spatially connected monitoring stations \citep{knight2020generalized}, economic indicators observed across trading countries \citep{nason2022quantifying}, and epidemiological outcomes recorded over interconnected regions \citep{meyer2017spatio}. In such settings, the evolution of a variable may depend not only on its own past but also on past observations at neighboring units. Incorporating this relational information can therefore be important for both parsimonious modeling and accurate forecasting.

A standard approach to multivariate time series modeling is the vector autoregressive (VAR) model, which allows each component series to depend on lagged values of all variables in the system. Although flexible, an unrestricted VAR of temporal order $p$ for $N$ variables contains $N^2p$ dynamic coefficients. Consequently, the number of parameters grows quadratically with the dimension of the system, which can lead to overfitting and poor out-of-sample performance, especially when the available time series is short relative to the number of variables. Bayesian vector autoregressive (BVAR) models address this problem by imposing structured shrinkage priors on the autoregressive coefficients. These priors help regularize highly parameterized VAR models and reduce overfitting \citep{banbura2010large,giannone2015prior}. In particular, Minnesota priors impose stronger shrinkage on more distant temporal lags \citep{repec:fip:fedmwp:115,litterman1986forecasting}, while dummy-observation priors such as the sum-of-coefficients (SOC) and dummy-initial-observation (DIO) priors incorporate prior information about persistence \citep{doan1984forecasting,sims1993nine, giannone2015prior,koop2010bayesian,kuschnig2021bvar}. 
More generally, the Bayesian framework provides posterior distributions for model parameters and future observations, allowing uncertainty quantification for both.

When the variables are associated with the nodes of a known network, however, an unrestricted VAR does not exploit this structural information. Autoregressive models for network time series include the network autoregressive (NAR) model of \citet{zhu2017network}, which incorporates own-lag and neighbor effects. A related model was proposed by \citet{knight2016modelling} and subsequently developed into the generalized network autoregressive (GNAR) framework by \citet{knight2020generalized}. GNAR allows higher-order temporal lags and neighborhoods, combining each node's own lagged observations with weighted averages of lagged observations from neighboring nodes. Bayesian approaches to autoregressive modeling of network time series have also been considered, including variational Bayesian inference for dynamic network autoregressive models \citep{lai2022variational} and, more recently, Bayesian mixture models for count-valued network autoregressive processes \citep{hung2025bayesian}.

In this paper, we combine the structural parsimony of GNAR with Bayesian shrinkage. We propose a Bayesian generalized network autoregressive (BGNAR) model by adapting key prior constructions from the BVAR literature to the GNAR framework. Specifically, we assign Minnesota-type shrinkage priors to the two types of dynamic coefficients in GNAR: the own-lag coefficients and the network-lag coefficients. For both coefficient types, the prior variance decreases with temporal lag, while the network-lag coefficients are additionally shrunk as neighborhood order increases. We further model the own-lag coefficients hierarchically at each temporal lag, allowing node-specific coefficients to share information across nodes while retaining heterogeneity. We also adapt the SOC and DIO priors to the GNAR parameterization to incorporate prior information on persistence. This framework allows a sufficiently rich model to be fitted without an exhaustive search over temporal and neighborhood orders, while providing posterior inference for both dynamic coefficients and future observations.


We evaluate the proposed method through simulation studies under different network topologies, dependence structures, and training-sample lengths, comparing BGNAR with frequentist GNAR and BVAR. The simulation studies examine forecasting performance, parameter recovery, and uncertainty quantification. We further illustrate the proposed method using the wind-speed network data considered by \citet{knight2020generalized}.

The remainder of the paper is organized as follows. Section~\ref{sec:preliminary} reviews the GNAR framework, its representation as a restricted VAR, and the BVAR prior constructions that motivate our approach. Section~\ref{sec:model} introduces the proposed BGNAR model, prior specification, and posterior computation. Section~\ref{sec:simul} presents the simulation studies, Section~\ref{sec:wind} applies the proposed method to the wind-speed network data, and Section~\ref{sec:conclusion} concludes.

\section{Preliminaries}\label{sec:preliminary}

\subsection{Network Notation}\label{subsec:net_not}
We begin by introducing the network structure and notation used throughout the paper. Let $\mathcal G = (\mathcal V, \mathcal E)$ be a fixed undirected graph with $N$ nodes, where $\mathcal V = \{1,\ldots,N\}$ denotes the node set and $\mathcal E$ denotes the edge set. The multivariate time series is observed on the nodes of $\mathcal G$, so that each component series is associated with one node of the network.

To describe dependence through the network, let $d_{\mathcal G}(i,q)$ denote the
shortest-path distance between nodes $i$ and $q$. For $r=1,2,\ldots$, the $r$th-order neighborhood of node $i$ is defined as
\[
\mathcal N_i^{(r)} = \left\{ q \in \mathcal V : d_{\mathcal G}(i,q) = r \right\}.
\]
Thus, $\mathcal N_i^{(1)}$ contains the immediate neighbors of node $i$, whereas
$\mathcal N_i^{(2)}$ contains nodes at graph distance two, and so forth.

For each neighborhood order $r$, define the $N \times N$ neighborhood-weight matrix
\begin{equation} \label{eq:w_r_mat}
    W^{(r)} = \left(w_{iq}^{(r)}\right)_{i,q=1}^N,
\end{equation}
where
\[
w_{iq}^{(r)} =
\begin{cases}
\dfrac{1}{|\mathcal N_i^{(r)}|},
& q \in \mathcal N_i^{(r)},\\[6pt]
0, & \text{otherwise}.
\end{cases}
\]
In particular, $w_{ii}^{(r)}=0$. For every nonempty $r$th-order neighborhood,
the weights therefore satisfy $\sum_{q=1}^N w_{iq}^{(r)} = 1$. Moreover,
\[
\left(W^{(r)}X_t\right)_i = \frac{1}{|\mathcal N_i^{(r)}|}
\sum_{q \in \mathcal N_i^{(r)}} X_{q,t},
\]
where $X_{q,t}$ denotes the observation at node $q$ and time $t$, so that this quantity represents the average over the $r$th-order neighbors of node $i$.

Let
\[
X_t = (X_{1,t},\ldots,X_{N,t})^\top \in \mathbb R^N, \qquad t=1,\ldots,T,
\]
denote the network time series observed on $\mathcal G$. Throughout the paper, $\operatorname{diag}(a)$ denotes the diagonal matrix with diagonal vector $a$,
$\operatorname{vec}(A)$ denotes the column-wise vectorization of a matrix $A$, and $\otimes$ denotes the Kronecker product. We write $\operatorname{blockdiag}(A_1,\ldots,A_k)$ for the block-diagonal matrix whose diagonal blocks are $A_1,\ldots,A_k$.

\subsection{Generalized Network Autoregressive Model}\label{subsec:gnar}
We review the generalized network autoregressive (GNAR) model of \citet{knight2020generalized}. Let $p$ denote the temporal order, that is, the maximum temporal lag included in the model. Thus,
$j=1,\ldots,p$ indexes temporal lag. For each temporal lag $j$, let $s_j$ denote the maximum neighborhood order included at that lag, so that $r=1,\ldots,s_j$ indexes neighborhood order. Writing $\bm s=[s_1,\ldots,s_p]$, the individual-$\alpha$ GNAR$(p,\bm s)$ model is
given by
\begin{equation} \label{eq:gnar_basic}
    X_{i,t}
=
\mu_i
+
\sum_{j=1}^p
\left[
\alpha_{i,j}X_{i,t-j}
+
\sum_{r=1}^{s_j}
\beta_{j,r}
\sum_{q\in\mathcal N_i^{(r)}}
w_{iq}^{(r)}X_{q,t-j}
\right]
+
u_{i,t},
\end{equation}
for $i=1,\ldots,N$ and $t=p+1,\ldots,T$. A more restrictive version is the global-$\alpha$ GNAR model, in which $\alpha_{i,j}=\alpha_j$ for all nodes $i$ at each temporal lag $j$.

Throughout the paper, $j$ denotes the temporal lag and $r$ denotes the neighborhood order. Accordingly, $\alpha_{i,j}$ is the own-lag coefficient for node $i$ at temporal lag $j$, whereas $\beta_{j,r}$ is the network-lag coefficient at temporal lag $j$ and neighborhood order $r$.

The GNAR model in (\ref{eq:gnar_basic}) can be written in vector form as
\begin{equation} \label{eq:gnar_fix1}
    X_t = \mu + \sum_{j=1}^p \Phi_j X_{t-j} + u_t,
\end{equation}
where $\mu = (\mu_1,\ldots,\mu_N)^\top$, $u_t = (u_{1,t},\ldots,u_{N,t})^\top \overset{\mathrm{i.i.d.}}{\sim} N_N(0,D_u)$ with $D_u = \operatorname{diag}(\sigma_1^2,\ldots,\sigma_N^2)$, and 
\[\Phi_j = \operatorname{diag}(\{\alpha_{i,j}\}_{i=1}^{N}) + \sum_{r=1}^{s_j} \beta_{j,r}W^{(r)}, \qquad j=1,\ldots,p.\]
For a fixed network, the GNAR model can be viewed as a restricted VAR model, where the restrictions on the coefficient matrices are determined by the network structure. 

The GNAR model in \eqref{eq:gnar_fix1} takes the linear regression form
\begin{equation}
X_t=H_t\gamma+u_t,
\qquad t=p+1,\ldots,T,
\label{eq:gnar-regression}
\end{equation}
where $\gamma=(\mu^\top,\theta_1^\top,\ldots,\theta_p^\top)^\top \in\mathbb R^{K_\gamma}$, $\theta_j=(\alpha_{1,j},\ldots,\alpha_{N,j},\beta_{j,1},\ldots,\beta_{j,s_j})^\top \in\mathbb R^{K_j}$, with $K_j=N+s_j$ and $K_\gamma=N+\sum_{j=1}^p K_j$. The design matrix $H_t\in\mathbb R^{N\times K_\gamma}$ depends only on the neighborhood-weight matrices and the observed lagged values $X_{t-1},\ldots,X_{t-p}$. Its explicit form is provided in Appendix~\ref{sec:H_t}. This representation is used to construct the Bayesian model in Section~\ref{sec:model}.

\subsection{Bayesian Vector Autoregressive Model}

Bayesian vector autoregressive (BVAR) models regularize the potentially large number of coefficients in an unrestricted VAR by incorporating prior information. Consider the VAR$(p)$ model
\[
X_t= \mu + \sum_{j=1}^p A_j X_{t-j} + u_t, \qquad u_t \sim N_N(0,\Sigma),
\]
where $\mu\in\mathbb R^N$ is an intercept vector, $A_j$ is an unrestricted $N\times N$ coefficient matrix at temporal lag $j$, and $\Sigma$ is the error covariance matrix. A BVAR model places prior distributions on these coefficient matrices, with the Minnesota prior and dummy-observation priors being widely used choices \citep{giannone2015prior,kuschnig2021bvar}.

The Minnesota prior imposes structured shrinkage on the VAR coefficients. In its conventional form, the first own-lag coefficient is centered around one, while the remaining coefficients are centered around zero:
\[
E\left[(A_j)_{iq}\right]
=
\begin{cases}
1, & i=q \text{ and } j=1,\\
0, & \text{otherwise}.
\end{cases}
\]
The prior variance typically decreases with the temporal lag. Schematically, the lag-decay structure can be written as
\[
\operatorname{Var}\left((A_j)_{iq}\mid\cdot\right)
\propto
\lambda^2 j^{-a},
\]
up to variable-specific scale adjustments, where $\lambda$ controls the overall amount of shrinkage and $a>0$ determines the rate of decay across temporal lags. Thus, coefficients at more distant lags are shrunk more strongly.

Prior information about persistent dynamics, where the current level of a series tends to carry over to future periods, can also be incorporated through dummy observations. Let $\bar X_0=p^{-1}\sum_{t=1}^p X_t$
denote the average of the initial $p$ observations. The sum-of-coefficients (SOC) prior favors
\[
\left(A_1+\cdots+A_p\right)\operatorname{diag}(\bar X_0)
\approx
\operatorname{diag}(\bar X_0).
\]
When the elements of $\bar X_0$ are nonzero, this corresponds to $A_1+\cdots+A_p\approx I_N$. Thus, the own-variable coefficients are encouraged to sum to one across lags, whereas the cross-variable coefficients are encouraged to sum to zero. Intuitively, this encourages each variable to retain its own level over time rather than being driven by persistent effects from other variables. The dummy-initial-observation (DIO) prior favors
\[
\mu + \left(A_1+\cdots+A_p\right)\bar X_0 \approx \bar X_0,
\]
encouraging the process to remain close to its initial level. Both priors can be incorporated by adding dummy observations to the model, with separate tightness parameters controlling the influence of the SOC and DIO restrictions.

The degree of shrinkage and the strength of the dummy-observation restrictions need not be fixed in advance. In a hierarchical Bayesian formulation, the hyperparameters controlling the shrinkage and the strength of the dummy-observation restrictions are assigned prior distributions and inferred jointly with the model parameters \citep{giannone2015prior,kuschnig2021bvar}.

The BGNAR model developed in the next section adapts these prior constructions to the GNAR framework. Specifically, the BGNAR model applies Minnesota-type temporal lag shrinkage to both the own-lag and network-lag coefficients and introduces additional shrinkage across neighborhood order for the network-lag coefficients. The SOC and DIO priors are likewise reformulated under the network-induced structure of the GNAR coefficient matrices.

\section{Bayesian Generalized Network Autoregressive Model}
\label{sec:model}
\color{black}
In this section, we introduce a prior specification for the BGNAR model, motivated by priors commonly used in BVAR, including the Minnesota, SOC, and DIO priors. 
\color{black}
Following the Minnesota prior in BVAR, we propose a Minnesota-type prior for BGNAR that incorporates shrinkage across temporal lags and neighborhood orders. We also adapt the SOC and DIO priors to BGNAR to incorporate prior information about persistence.

Recalling the notation introduced in Section~\ref{subsec:gnar}, we use the regression representation in \eqref{eq:gnar-regression} to specify the BGNAR model as
\[
X_t \mid X_{t-1},\ldots,X_{t-p},\gamma,D_u
\sim N_N(H_t\gamma,D_u),
\qquad t=p+1,\ldots,T.
\]
We assume that the error covariance matrix is diagonal, which rules out contemporaneous correlations between errors at different nodes and avoids the need to estimate a full
$N\times N$ covariance matrix. We assign independent inverse-gamma priors to the error variances:
\[
\sigma_i^2 \overset{\mathrm{i.i.d.}}{\sim} \mathrm{IG}(c_\sigma,d_\sigma),
\qquad i=1,\ldots,N,
\]
where $\mathrm{IG}(c,d)$ denotes the inverse-gamma distribution with shape parameter $c>0$ and scale parameter $d>0$.
For the intercept vector $\mu$, we assign the Gaussian prior
\[
\mu \sim N_N(0,c_\mu I_N).
\]

For ease of reference, Table~\ref{tab:notation} summarizes the main notation used in the prior specification and posterior computation below.

\begin{table}[!t]
\centering
\caption{Main notation for the BGNAR model.}
\label{tab:notation}
\small
\begin{tabular}{ll}
\toprule
Symbol & Description \\
\midrule
$\mu$ & $N$-dimensional intercept vector \\
$\alpha_{i,j}$ & own-lag coefficient for node $i$ at temporal lag $j$ \\
$\beta_{j,r}$ & network-lag coefficient at temporal lag $j$ and neighborhood order $r$ \\
$\theta_j$ & vector of own-lag and network-lag coefficients at temporal lag $j$ 
\\
$\gamma$ & regression coefficient vector collecting $\mu$ and $\theta_1,\ldots,\theta_p$ \\
$b_j$ & common component of the own-lag coefficients at temporal lag $j$ \\
$\kappa_\alpha$ & own-lag heterogeneity parameter \\
$\lambda_\alpha^2$ & overall prior variance scale for the own-lag coefficients \\
$\lambda_\beta^2$ & overall prior variance scale for the network-lag coefficients \\
$a$ & temporal-lag decay parameter \\
$d$ & neighborhood-order decay parameter \\
$\sigma_i^2$ & error variance for node $i$ \\
$D_u$ & diagonal error covariance matrix \\
$\tau_{\mathrm{soc}}^2$ & SOC dummy-prior tightness parameter \\
$\tau_{\mathrm{dio}}^2$ & DIO dummy-prior tightness parameter \\
\bottomrule
\end{tabular}
\end{table}

\subsection{Minnesota-Type Prior}
We propose a Minnesota-type prior for the own-lag coefficients $\alpha$ and the network-lag coefficients $\beta$ defined in Section~\ref{subsec:gnar}. For $\alpha$, the prior variances decrease with temporal lag. For $\beta$, the prior variances decrease with both temporal lag and neighborhood order.

\paragraph{Prior of $\alpha$}
For the own-lag coefficients, we use the following reparameterization:
\[
\alpha_{i,j} = b_j + \delta_{i,j},
\]
where $b_j$ represents the common component at temporal lag $j$, and $\delta_{i,j}$ represents the node-specific deviation from $b_j$. We assign the following conditionally independent Gaussian priors:
\begin{align*}
    b_j \mid \lambda_\alpha^2,a
    &\sim N(0,\lambda_\alpha^2 j^{-a}),\\
    \delta_{i,j} \mid \kappa_\alpha,\lambda_\alpha^2,a
    &\sim N(0,\kappa_\alpha\lambda_\alpha^2 j^{-a}),
\end{align*}
where $a>0$ controls the rate of prior variance decay across temporal lags, $\lambda_\alpha^2$ determines the overall prior variance scale for the own-lag coefficients, and $\kappa_\alpha>0$ controls the degree of node-specific variation around the common component. The factor $j^{-a}$ induces stronger shrinkage toward zero as the temporal lag $j$ increases.

For the temporal-lag decay parameter $a$, we assign a discrete uniform prior on $\mathcal A=\{1,1.5,2,2.5,3\}$. This grid allows for different degrees of shrinkage across temporal lags and permits a categorical update of $a$ within Gibbs sampling. The same parameter $a$ is used for the network-lag coefficients $\beta$.

Finally, we assign inverse-gamma priors to
$\lambda_\alpha^2$ and $\kappa_\alpha$:
\[
\lambda_\alpha^2 \sim \mathrm{IG}(c_\alpha,d_\alpha),
\qquad
\kappa_\alpha \sim \mathrm{IG}(c_\kappa,d_\kappa).
\]

\paragraph{Prior of $\beta$}
For each temporal lag $j$, let $\beta_j=(\beta_{j,1},\ldots,\beta_{j,s_j})^\top$ denote the vector of network-lag coefficients. We assign the following conditionally independent Gaussian priors:
\[
\beta_j\mid \lambda_\beta^2,a,d
\overset{\mathrm{ind}}{\sim}
N_{s_j}(0,\lambda_\beta^2 j^{-a}D_{\beta,j}(d)),
\qquad j=1,\ldots,p,
\]
where $\lambda_\beta^2$ controls the overall prior variance scale for the network-lag coefficients. To incorporate additional shrinkage across neighborhood orders, we define
\[
D_{\beta,j}(d) = \operatorname{diag}(1^{-d},2^{-d},\ldots,s_j^{-d}),
\]
where $d\in\mathcal D=\{1,1.5,2,2.5,3\}$. The hyperparameter $d$ controls the rate of prior variance decay across neighborhood orders. As with $a$, we assign a discrete uniform prior to $d$ on $\mathcal D$. We assign an inverse-gamma prior to
$\lambda_\beta^2$:
\[
\lambda_\beta^2 \sim \mathrm{IG}(c_\beta,d_\beta).
\]

In particular, the conditional prior variance of each network-lag coefficient is given by
\[
\operatorname{Var}(\beta_{j,r}\mid\lambda_\beta^2,a,d) = \lambda_\beta^2 j^{-a}r^{-d}.
\]
Thus, for the network-lag coefficients $\beta$, $a$ governs the decay across temporal lags, while $d$ governs the decay across neighborhood orders.

\paragraph{Prior of $\theta_j$}
Combining the priors for $\alpha$ and $\beta$, the prior for $\theta_j$ defined in Section~\ref{subsec:gnar} can be written as
\[
\theta_j \mid
b_j,\kappa_\alpha,\lambda_\alpha^2,
\lambda_\beta^2,a,d
\overset{\mathrm{ind}}{\sim}
N_{K_j}(m_{\theta,j},V_{\theta,j}),
\qquad j=1,\ldots,p,
\]
where \[
m_{\theta,j}
=
\begin{pmatrix}
    b_j 1_N \\
    0_{s_j}
\end{pmatrix},
\qquad
V_{\theta,j}
=
\begin{pmatrix}
    \kappa_\alpha \lambda_\alpha^2 j^{-a} I_N & 0 \\
    0 & \lambda_\beta^2 j^{-a} D_{\beta,j}(d)
\end{pmatrix}.
\]

\subsection{Dummy-Observation Priors}\label{subsec:dummy-priors}
Following \cite{giannone2015prior}, we incorporate the sum-of-coefficients (SOC) and dummy-initial-observation (DIO) priors into the BGNAR model through dummy observations. These priors encode prior information about persistence.

To incorporate these priors, we add dummy observations with response vectors $x_{\mathrm{soc}}^\ast$ and $x_{\mathrm{dio}}$ and design matrices $H_{\mathrm{soc}}^\ast$ and $H_{\mathrm{dio}}$ to the regression model in \eqref{eq:gnar-regression}.
First, we stack the observed responses and design matrices as
\[
x_{\mathrm{o}} = \begin{pmatrix}
X_T\\X_{T-1}\\\vdots\\X_{p+1}
\end{pmatrix}
\in\mathbb{R}^{NT_\ast},
\qquad
H_{\mathrm{o}} = \begin{pmatrix}
H_T\\H_{T-1}\\\vdots\\H_{p+1}
\end{pmatrix}
\in\mathbb{R}^{NT_\ast\times K_\gamma},
\]
where $T_\ast=T-p$. The resulting regression model is
\[
x_{\mathrm{o}}=H_{\mathrm{o}}\gamma+\epsilon_{\mathrm{o}},
\qquad
\epsilon_{\mathrm{o}}\sim N_{NT_\ast}(0,I_{T_\ast}\otimes D_u).
\]

By incorporating the dummy observations, we obtain the following augmented regression model:
\[
x_\ast=H_\ast\gamma+\epsilon_\ast,
\]
where
\[
x_\ast
=
\begin{pmatrix}
x_{\mathrm{o}}\\x_{\mathrm{dio}}\\x_{\mathrm{soc}}^\ast
\end{pmatrix},
\qquad
H_\ast
=
\begin{pmatrix}
H_{\mathrm{o}}\\H_{\mathrm{dio}}\\H_{\mathrm{soc}}^\ast
\end{pmatrix},
\qquad
\epsilon_\ast
=
\begin{pmatrix}
\epsilon_{\mathrm{o}}\\\epsilon_{\mathrm{dio}}\\\epsilon_{\mathrm{soc}}^\ast
\end{pmatrix},
\]
and the SOC and DIO dummy response vectors, design matrices, and error terms
are defined below.

\paragraph{Sum-of-Coefficients Prior}
The SOC prior favors
\begin{align}
    \label{eq:target_soc}
    (\Phi_1+\cdots+\Phi_p)\operatorname{diag}(\bar X_0)
    \approx \operatorname{diag}(\bar X_0),
\end{align}
where $\bar X_0=p^{-1}\sum_{t=1}^p X_t$. To incorporate this relationship through dummy observations, we define
\[
X_{\mathrm{soc}}
=
\operatorname{diag}(\bar X_0)
\in\mathbb R^{N\times N},
\qquad
Y_{\mathrm{soc}}
=
\begin{pmatrix}
0_N^\top\\
X_{\mathrm{soc}}\\
\vdots\\
X_{\mathrm{soc}}
\end{pmatrix}
\in\mathbb R^{(1+Np)\times N},
\]
where $Y_{\mathrm{soc}}$ contains $p$ copies of $X_{\mathrm{soc}}$. The target relationship in \eqref{eq:target_soc} can then be written as
\[
X_{\mathrm{soc}}\approx BY_{\mathrm{soc}},\qquad
B=[\mu,\Phi_1,\ldots,\Phi_p]
\in\mathbb R^{N\times(1+Np)}.
\]


Since $BY_{\mathrm{soc}}$ is linear in $\gamma$, we define the design matrix $H_{\mathrm{soc}}\in\mathbb R^{N^2\times K_\gamma}$ such that
\[
H_{\mathrm{soc}}\gamma =
\operatorname{vec}(BY_{\mathrm{soc}}).
\]
Its explicit construction is provided in Appendix~\ref{sec:soc_dio}.
Letting $x_{\mathrm{soc}}=\operatorname{vec}(X_{\mathrm{soc}})
\in\mathbb R^{N^2}$, we obtain the Gaussian pseudo-observation model
\[
x_{\mathrm{soc}}
=
H_{\mathrm{soc}}\gamma+\epsilon_{\mathrm{soc}},
\qquad
\epsilon_{\mathrm{soc}}\mid\tau_{\mathrm{soc}}^2
\sim N_{N^2}(0,\tau_{\mathrm{soc}}^2 I_{N^2}),
\]
where $\tau_{\mathrm{soc}}^2$ is a tightness parameter controlling the strength of the SOC prior. Smaller values of $\tau_{\mathrm{soc}}^2$ impose stronger shrinkage toward the target relationship.

\paragraph{Reduced SOC Block}

The SOC dummy-observation introduces $N^2$ dummy observations, which can lead to an overly concentrated full conditional distribution for $\tau_{\mathrm{soc}}^2$. To address this issue, we construct a reduced SOC block that preserves the residual sum of squares.

Let
\[
M_{\mathrm{soc}}
=
[x_{\mathrm{soc}},H_{\mathrm{soc}}]
\in\mathbb R^{N^2\times(1+K_\gamma)},
\]
and consider its compact singular value decomposition
\[
M_{\mathrm{soc}}=U_qD_qV_q^\top,
\]
where $q=\operatorname{rank}(M_{\mathrm{soc}})$, $U_q\in\mathbb R^{N^2\times q}$, $D_q\in\mathbb R^{q\times q}$, and $V_q\in\mathbb R^{(1+K_\gamma)\times q}$, with $U_q^\top U_q=V_q^\top V_q=I_q$.
We define the reduced SOC block as
\[
[x_{\mathrm{soc}}^\ast,H_{\mathrm{soc}}^\ast]
=
U_q^\top M_{\mathrm{soc}}
=
D_qV_q^\top.
\]
Since $x_{\mathrm{soc}}-H_{\mathrm{soc}}\gamma$ belongs to the column space of $M_{\mathrm{soc}}$, we have
\[
\|x_{\mathrm{soc}}-H_{\mathrm{soc}}\gamma\|^2
=
\|x_{\mathrm{soc}}^\ast-H_{\mathrm{soc}}^\ast\gamma\|^2.
\]
Thus, the reduced block preserves the SOC residual penalty while requiring only $q$ dummy observations.

We use the reduced SOC pseudo-observation model
\[
x_{\mathrm{soc}}^\ast
=
H_{\mathrm{soc}}^\ast\gamma+\epsilon_{\mathrm{soc}}^\ast,
\qquad
\epsilon_{\mathrm{soc}}^\ast\mid\tau_{\mathrm{soc}}^2
\sim N_q(0,\tau_{\mathrm{soc}}^2I_q),
\]
and assign an inverse-gamma prior to
$\tau_{\mathrm{soc}}^2$:
\[
\tau_{\mathrm{soc}}^2\sim \mathrm{IG}(c_{\mathrm{soc}},d_{\mathrm{soc}}).
\]

\paragraph{Dummy-Initial-Observation Prior}
The DIO prior targets the following relationship:
\begin{align}
    \label{eq:target_dio}
    \mu + (\Phi_1+\cdots+\Phi_p)\bar X_0
    \approx \bar X_0.
\end{align}
This prior complements the SOC prior by allowing for possible cointegration relationships among variables. It encourages the conditional mean to approximately preserve the initial average level of the time series.

To incorporate this relationship through dummy observations, we define
\[
X_{\mathrm{dio}}
=
\bar X_0\in\mathbb R^{N\times 1},
\qquad
Y_{\mathrm{dio}}
=
\begin{pmatrix}
1\\
\bar X_0\\
\vdots\\
\bar X_0
\end{pmatrix}
\in\mathbb R^{(1+Np)\times 1},
\]
where $Y_{\mathrm{dio}}$ contains $p$ copies of $\bar X_0$. The target relationship in \eqref{eq:target_dio} can then be written as
\[
X_{\mathrm{dio}}\approx BY_{\mathrm{dio}}.
\]
Let $x_{\mathrm{dio}}=X_{\mathrm{dio}}$ and define the design matrix
$H_{\mathrm{dio}}\in\mathbb R^{N\times K_\gamma}$
such that
\[
H_{\mathrm{dio}}\gamma=BY_{\mathrm{dio}}.
\]
The explicit construction of $H_{\mathrm{dio}}$ is provided in  Appendix~\ref{sec:soc_dio}.

The corresponding Gaussian pseudo-observation model is
\[
x_{\mathrm{dio}}
=
H_{\mathrm{dio}}\gamma+\epsilon_{\mathrm{dio}},
\qquad
\epsilon_{\mathrm{dio}}\mid\tau_{\mathrm{dio}}^2
\sim N_N(0,\tau_{\mathrm{dio}}^2I_N),
\]
where $\tau_{\mathrm{dio}}^2$ is a tightness parameter controlling the strength of the DIO prior.
Smaller values of $\tau_{\mathrm{dio}}^2$ impose stronger shrinkage toward the target relationship.
We assign an inverse-gamma prior to $\tau_{\mathrm{dio}}^2$:
\[
\tau_{\mathrm{dio}}^2\sim \mathrm{IG}(c_{\mathrm{dio}},d_{\mathrm{dio}}).
\]

\subsection{Posterior Computation}\label{sec:posterior}
With $b=(b_1,\ldots,b_p)^\top$, the Minnesota-type prior and the intercept prior imply the following prior $\gamma$:
\[
\gamma\mid b,\kappa_\alpha,\lambda_\alpha^2,
\lambda_\beta^2,a,d
\sim N_{K_\gamma}(m_\gamma(b),V_\gamma),
\]
where
\[
m_\gamma(b)
=
(0_N^\top,m_{\theta,1}^\top,\ldots,
m_{\theta,p}^\top)^\top, \qquad 
V_\gamma
=
\operatorname{blockdiag}
(c_\mu I_N,V_{\theta,1},\ldots,V_{\theta,p}).
\]
Together with the observed-data likelihood and the two dummy-observation blocks, the prior hierarchy yields the joint posterior density
\begin{align*}
&\pi\left(
\gamma,b,\kappa_\alpha,D_u,
\lambda_\alpha^2,\lambda_\beta^2,a,d,
\tau_{\mathrm{soc}}^2,\tau_{\mathrm{dio}}^2
\mid\{X_t\}_{t=1}^T
\right)
\\
&\quad\propto
\left[
\prod_{t=p+1}^T
\phi_N(X_t;H_t\gamma,D_u)
\right]
\phi_q\left(
x_{\mathrm{soc}}^\ast;
H_{\mathrm{soc}}^\ast\gamma,
\tau_{\mathrm{soc}}^2I_q
\right)
\\
&\qquad\times
\phi_N\left(
x_{\mathrm{dio}};
H_{\mathrm{dio}}\gamma,
\tau_{\mathrm{dio}}^2I_N
\right)
\phi_{K_\gamma}(\gamma;m_\gamma(b),V_\gamma)
\\
&\qquad\times
\prod_{j=1}^p
\phi_1(b_j;0,\lambda_\alpha^2j^{-a})
\times
\mathrm{IG}(\kappa_\alpha;c_\kappa,d_\kappa)
\prod_{i=1}^N
\mathrm{IG}(\sigma_i^2;c_\sigma,d_\sigma)
\\
&\qquad\times
\mathrm{IG}(\lambda_\alpha^2;c_\alpha,d_\alpha)
\mathrm{IG}(\lambda_\beta^2;c_\beta,d_\beta)
\pi_a(a)\pi_d(d)
\\
&\qquad\times
\mathrm{IG}(\tau_{\mathrm{soc}}^2;c_{\mathrm{soc}},d_{\mathrm{soc}})
\mathrm{IG}(\tau_{\mathrm{dio}}^2;c_{\mathrm{dio}},d_{\mathrm{dio}}),
\end{align*}
where $\phi_k(\cdot;m,V)$ denotes the density of a $k$-variate Gaussian distribution, and $\pi_a$ and $\pi_d$ denote the prior probability mass functions of $a$ and $d$, respectively. All full conditional distributions are Gaussian, inverse-gamma, or categorical, allowing posterior sampling by a Gibbs sampler. At each iteration, the sampler updates the regression coefficients $\gamma$, the error variances, the common own-lag effects $b_j$, the shrinkage and heterogeneity parameters, the temporal- and neighborhood-decay parameters, and the SOC and DIO tightness parameters. 

The explicit full conditional distributions and Gibbs sampling steps are given in Appendix~\ref{app:gibbs}. \color{black} The hyperparameter settings used in our numerical experiments are provided in Appendix~\ref{sec:hyperparameter}. \color{black} The code for the BGNAR Gibbs sampler is available at \url{https://github.com/zlatjdals/BGNAR}.



\section{Simulation Studies} \label{sec:simul}

This section evaluates the forecasting performance, parameter estimation, and uncertainty quantification of the proposed BGNAR model. 
We consider three network designs: (i) Erd\H{o}s--R\'enyi (ER; \citealp{erdds1959random}), (ii) stochastic block model (SBM; \citealp{holland1983stochastic}), and (iii) small-world network (SWN; \citealp{watts1998collective}). For the ER design, each undirected edge is present independently with probability $0.2$. For the SBM design, the nodes are divided into four equal-sized blocks, with edge probabilities $0.7$ within blocks and $0.075$ between distinct blocks. For the SWN design, we begin with a one-dimensional ring lattice in which each node is connected to its two nearest neighbors on either side, and rewire each edge with probability $0.1$.

For the data-generating processes (DGPs), we consider the following
global-$\alpha$ GNAR model:
\[
X_t
=
0.1 \cdot 1_N
+
\sum_{j=1}^{p}
\left\{
\alpha_j I_N
+
\sum_{r=1}^{s_j}\beta_{j,r}W^{(r)}
\right\}X_{t-j}
+
\epsilon_t,
\qquad
\epsilon_t\sim N_N(0,0.1 I_N).
\]
Table~\ref{tab:dgp-design} summarizes the four DGPs. The first three DGPs share the same first-order structure, $(p,\bm s)=(1,[1])$, with $\alpha_1+\beta_{1,1}=0.9$. They differ in the relative strength of the own-lag and network-lag effects: $\mathrm{M}_{11}$ is own-lag dominant, $\mathrm{M}_{12}$ gives equal weight to the two effects, and $\mathrm{M}_{13}$ is network-lag dominant. $\mathrm{M}_{2}$ represents a higher-order setting with $(p,\bm s)=(2,[2,1])$.

\begin{table}[!t]
\centering
\caption{Four data-generating processes (DGPs) under the global-$\alpha$ GNAR model}
\label{tab:dgp-design}
\small
\begin{tabular}{lccc}
\toprule
DGP & True order & Own-lag coefficients & Network-lag coefficients  \\
\midrule
$\mathrm{M}_{11}$ & $(1,[1])$ & $\alpha_1=0.70$ & $\beta_{1,1}=0.20$  \\
$\mathrm{M}_{12}$ & $(1,[1])$ & $\alpha_1=0.45$ & $\beta_{1,1}=0.45$  \\
$\mathrm{M}_{13}$ & $(1,[1])$ & $\alpha_1=0.20$ & $\beta_{1,1}=0.70$  \\
$\mathrm{M}_{2}$  & $(2,[2,1])$ & $(\alpha_1,\alpha_2)=(0.30,0.20)$ &
$(\beta_{1,1},\beta_{1,2})=(0.20,0.10),\ \beta_{2,1}=0.10$  \\
\bottomrule
\end{tabular}
\end{table}

For each network design and DGP, we consider $N=20$ nodes and sample sizes $T\in\{20,40,80,160\}$, with 100 replications for each setting. BGNAR is fitted with individual own-lag coefficients and a fixed over-specified structure $(p,\bm s)=(5,[3,3,3,3,3])$. We compare BGNAR with GNAR and BVAR. For GNAR, we center the time series and use the Bayesian information criterion (BIC) to select the temporal order, neighborhood orders, and global- or individual-$\alpha$ specification, with maximum temporal and neighborhood orders of five and three, respectively. For BVAR, we fix the temporal order at $p=5$. GNAR and BVAR are fitted using the R packages \texttt{GNAR} and \texttt{BVAR}, respectively.

\begin{figure}[!t]
\centering
\includegraphics[width=\textwidth]{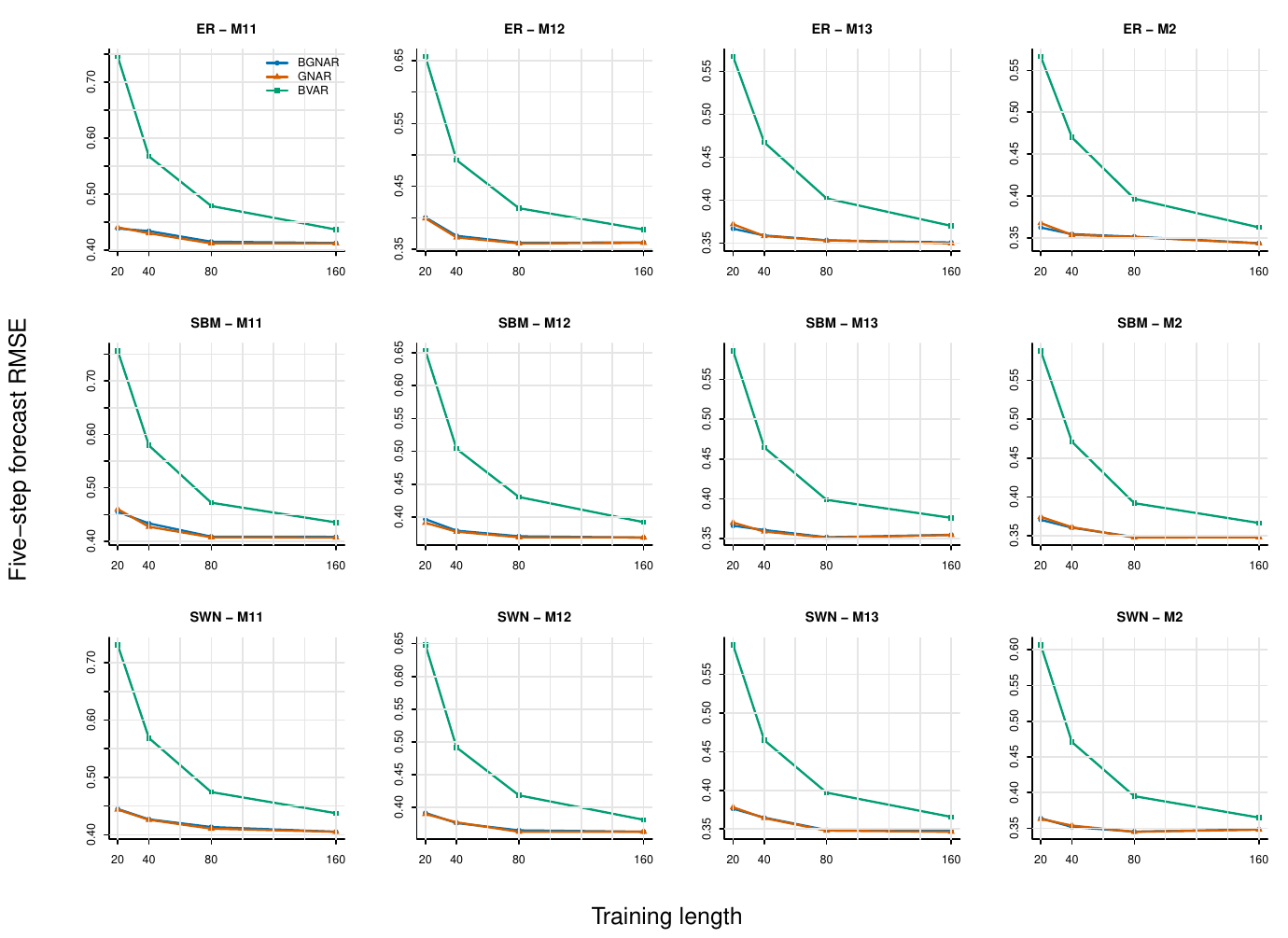}
\caption{Five-step forecast RMSE across the three network designs (ER, SBM, and SWN), four DGPs ($\mathrm{M}_{11}$, $\mathrm{M}_{12}$, $\mathrm{M}_{13}$, and $\mathrm{M}_{2}$), and training sample sizes.}
\label{fig:simulation-forecast}
\end{figure}

\begin{table}[!t]
\centering
\caption{In-sample fitting RMSE, five-step forecast RMSE, and coverage and average width of 95\% posterior predictive intervals across sample sizes and network designs. Values in parentheses are standard deviations of RMSE. Predictive intervals are reported only for BGNAR and BVAR.}
\label{tab:performance}
\footnotesize
\setlength{\tabcolsep}{5pt}
\begin{tabular}{clrrrr}
\toprule
$T$ & Method & Fit RMSE & Forecast RMSE & Coverage & Width \\
\midrule

\multicolumn{6}{l}{\textit{Erd\H{o}s--R\'enyi (ER)}} \\
20  & BGNAR & 0.290 (0.013) & \textbf{0.392 (0.055)} & 0.938 & 1.536 \\
    & GNAR  & 0.299 (0.013) & 0.395 (0.057)          & NA    & NA    \\
    & BVAR  & \textbf{0.080 (0.015)} & 0.634 (0.167) & 0.989 & 7.882 \\
\addlinespace
40  & BGNAR & 0.304 (0.008) & 0.380 (0.050)          & 0.940 & 1.479 \\
    & GNAR  & 0.310 (0.008) & \textbf{0.378 (0.050)} & NA    & NA    \\
    & BVAR  & \textbf{0.111 (0.013)} & 0.499 (0.088) & 0.973 & 2.949 \\
\addlinespace
80  & BGNAR & 0.310 (0.006) & 0.370 (0.046)          & 0.944 & 1.449 \\
    & GNAR  & 0.313 (0.006) & \textbf{0.369 (0.045)} & NA    & NA    \\
    & BVAR  & \textbf{0.205 (0.022)} & 0.423 (0.062) & 0.871 & 1.366 \\
\addlinespace
160 & BGNAR & 0.313 (0.004) & 0.367 (0.045)          & 0.947 & 1.435 \\
    & GNAR  & 0.315 (0.004) & \textbf{0.366 (0.045)} & NA    & NA    \\
    & BVAR  & \textbf{0.269 (0.012)} & 0.388 (0.049) & 0.739 & 0.899 \\

\midrule
\multicolumn{6}{l}{\textit{Stochastic block model (SBM)}} \\
20  & BGNAR & 0.289 (0.013) & \textbf{0.397 (0.058)} & 0.933 & 1.522 \\
    & GNAR  & 0.297 (0.013) & 0.399 (0.063)          & NA    & NA    \\
    & BVAR  & \textbf{0.079 (0.014)} & 0.646 (0.167) & 0.988 & 7.985 \\
\addlinespace
40  & BGNAR & 0.303 (0.009) & 0.384 (0.051)          & 0.937 & 1.470 \\
    & GNAR  & 0.309 (0.009) & \textbf{0.381 (0.050)} & NA    & NA    \\
    & BVAR  & \textbf{0.111 (0.012)} & 0.505 (0.106) & 0.971 & 2.919 \\
\addlinespace
80  & BGNAR & 0.309 (0.006) & 0.370 (0.044)          & 0.944 & 1.443 \\
    & GNAR  & 0.313 (0.006) & \textbf{0.369 (0.043)} & NA    & NA    \\
    & BVAR  & \textbf{0.205 (0.022)} & 0.423 (0.060) & 0.867 & 1.344 \\
\addlinespace
160 & BGNAR & 0.313 (0.004) & \textbf{0.370 (0.044)} & 0.944 & 1.431 \\
    & GNAR  & 0.314 (0.004) & \textbf{0.370 (0.045)} & NA    & NA    \\
    & BVAR  & \textbf{0.268 (0.012)} & 0.393 (0.050) & 0.736 & 0.900 \\

\midrule
\multicolumn{6}{l}{\textit{Small-world network (SWN)}} \\
20  & BGNAR & 0.290 (0.013) & 0.394 (0.053)          & 0.932 & 1.522 \\
    & GNAR  & 0.298 (0.014) & \textbf{0.394 (0.056)} & NA    & NA    \\
    & BVAR  & \textbf{0.080 (0.013)} & 0.643 (0.153) & 0.988 & 7.878 \\
\addlinespace
40  & BGNAR & 0.304 (0.009) & \textbf{0.380 (0.048)} & 0.940 & 1.469 \\
    & GNAR  & 0.310 (0.009) & 0.380 (0.048)          & NA    & NA    \\
    & BVAR  & \textbf{0.110 (0.012)} & 0.499 (0.099) & 0.970 & 2.939 \\
\addlinespace
80  & BGNAR & 0.311 (0.006) & 0.368 (0.046)          & 0.944 & 1.443 \\
    & GNAR  & 0.314 (0.006) & \textbf{0.367 (0.045)} & NA    & NA    \\
    & BVAR  & \textbf{0.207 (0.022)} & 0.421 (0.061) & 0.873 & 1.366 \\
\addlinespace
160 & BGNAR & 0.313 (0.004) & 0.366 (0.042)          & 0.945 & 1.427 \\
    & GNAR  & 0.315 (0.004) & \textbf{0.366 (0.042)} & NA    & NA    \\
    & BVAR  & \textbf{0.268 (0.012)} & 0.387 (0.050) & 0.735 & 0.882 \\

\bottomrule
\end{tabular}
\end{table}

To evaluate forecasting performance, we calculate the forecast root mean squared error (RMSE) as
\[
\mathrm{RMSE}_{\mathrm{forecast}}
=
\left[
\frac{1}{HN}
\sum_{h=1}^{H}
\sum_{i=1}^{N}
(\hat X_{i,T+h}-X_{i,T+h})^2
\right]^{1/2},
\]
where $X_{i,T+h}$ and $\hat{X}_{i,T+h}$ denote the realized value and the point forecast, respectively, for node $i$ at horizon $h$. For BGNAR, point forecasts are given by the posterior predictive means. We set the forecast horizon to $H=5$ in all settings. We also evaluate the in-sample fitting performance using
\[
\mathrm{RMSE}_{\mathrm{fitting}}
=
\left[
\frac{1}{N(T-5)}
\sum_{t=6}^{T}
\sum_{i=1}^{N}
(\tilde X_{i,t}-X_{i,t})^2
\right]^{1/2},
\]
where $\tilde X_{i,t}$ denotes the in-sample fitted value for node $i$ at time $t$. For a consistent comparison across methods, we evaluate the fitting RMSE over $t=6,\ldots,T$, corresponding to the maximum temporal order of five.

Figure~\ref{fig:simulation-forecast} presents the forecast RMSE for each network design and DGP. At $T=20$, BGNAR has slightly smaller forecast RMSE than GNAR under the ER and SBM networks, while the two methods perform similarly under the SWN. As $T$ increases, BGNAR and GNAR remain close across all three network designs. In contrast, BVAR consistently achieves the smallest fitting RMSE but the largest forecast RMSE, as shown in Table~\ref{tab:performance}, indicating that its better in-sample fit does not translate into improved forecasting performance.

Table~\ref{tab:performance} also reports the empirical coverage and average width of 95\% posterior predictive intervals for BGNAR and BVAR. For $T=20$ and $40$, BVAR has higher coverage but substantially wider intervals. As $T$ increases, its intervals become narrower and the coverage decreases markedly, falling below $0.75$ at $T=160$. In contrast, BGNAR maintains relatively stable interval widths and coverage close to the nominal level of $0.95$ across all sample sizes and network designs; even at $T=20$, its coverage exceeds $0.93$ in all three networks. These results indicate that BGNAR provides more stable predictive uncertainty quantification across different sample sizes.

\begin{table}[!t]
\centering
\caption{Parameter estimation RMSE for BGNAR and GNAR. Active and inactive RMSEs are computed over coefficients that are nonzero and zero in the DGPs, respectively. Full RMSE is computed over all coefficients. Values in parentheses denote standard deviations.}
\label{tab:parameter-recovery}
\footnotesize
\setlength{\tabcolsep}{4.2pt}
\begin{tabular}{cclrrr}
\toprule
$T$ & Parameter & Method & Active RMSE & Inactive RMSE & Full RMSE \\
\midrule
20 & $\alpha$ & BGNAR & 0.115 (0.047) & \textbf{0.028 (0.013)} & \textbf{0.063 (0.021)} \\
   &           & GNAR  & \textbf{0.085 (0.052)} & 0.074 (0.028) & 0.081 (0.026) \\
   & $\beta$  & BGNAR & 0.122 (0.080) & \textbf{0.026 (0.013)} & \textbf{0.046 (0.019)} \\
   &           & GNAR  & \textbf{0.101 (0.069)} & 0.071 (0.045) & 0.080 (0.039) \\
\addlinespace
40 & $\alpha$ & BGNAR & 0.078 (0.024) & \textbf{0.023 (0.010)} & 0.044 (0.011) \\
   &           & GNAR  & \textbf{0.040 (0.027)} & 0.043 (0.015) & \textbf{0.044 (0.013)} \\
   & $\beta$  & BGNAR & 0.070 (0.047) & 0.023 (0.009) & \textbf{0.032 (0.011)} \\
   &           & GNAR  & \textbf{0.068 (0.048)} & \textbf{0.018 (0.027)} & 0.034 (0.023) \\
\addlinespace
80 & $\alpha$ & BGNAR & 0.058 (0.015) & \textbf{0.018 (0.007)} & 0.033 (0.007) \\
   &           & GNAR  & \textbf{0.024 (0.017)} & 0.027 (0.010) & \textbf{0.027 (0.009)} \\
   & $\beta$  & BGNAR & 0.043 (0.030) & 0.019 (0.007) & 0.024 (0.008) \\
   &           & GNAR  & \textbf{0.043 (0.030)} & \textbf{0.005 (0.013)} & \textbf{0.018 (0.015)} \\
\addlinespace
160 & $\alpha$ & BGNAR & 0.043 (0.009) & \textbf{0.015 (0.005)} & 0.025 (0.005) \\
   &           & GNAR  & \textbf{0.016 (0.011)} & 0.018 (0.007) & \textbf{0.018 (0.006)} \\
   & $\beta$  & BGNAR & 0.029 (0.019) & 0.015 (0.005) & 0.018 (0.006) \\
   &           & GNAR  & \textbf{0.028 (0.021)} & \textbf{0.002 (0.007)} & \textbf{0.011 (0.010)} \\
\bottomrule
\end{tabular}
\end{table}

Table~\ref{tab:parameter-recovery} presents the parameter estimation performance of BGNAR and GNAR. Active RMSE measures the error for nonzero coefficients in the DGPs, inactive RMSE measures the error for zero coefficients within the fixed over-specified structure $(p,\bm s)=(5,[3,3,3,3,3])$, and full RMSE is computed over all coefficients. BVAR is excluded because it does not use the GNAR coefficient structure. At $T=20$, BGNAR has smaller inactive and full RMSE for both $\alpha$ and $\beta$, whereas GNAR has smaller active RMSE. For $\alpha$, BGNAR retains smaller inactive RMSE across all sample sizes. For $\beta$, this advantage is limited to $T=20$, while GNAR performs better for larger $T$. The larger active RMSE for BGNAR is consistent with the shrinkage imposed on nonzero coefficients. In contrast, BGNAR estimates inactive coefficients more accurately, particularly in short samples, which helps regularize the fixed over-specified model.

Overall, BGNAR achieves forecasting performance comparable to GNAR and better than BVAR across the simulation settings considered, despite using a fixed over-specified lag structure without explicit order selection. BGNAR also more strongly regularizes inactive coefficients, particularly in short samples, although estimation of active coefficients can be less accurate than under GNAR. Moreover, its posterior predictive intervals maintain coverage close to the nominal level across sample sizes, whereas BVAR exhibits greater variability in coverage and interval width.

Additional results on the empirical coverage and average width of the 95\% credible intervals for the BGNAR coefficients are reported in Appendix~\ref{sec:additional_simul}.

\color{black}
\section{Wind-Speed Network Application}\label{sec:wind}
\begin{figure}[!t]
\centering
\includegraphics[width=\textwidth]{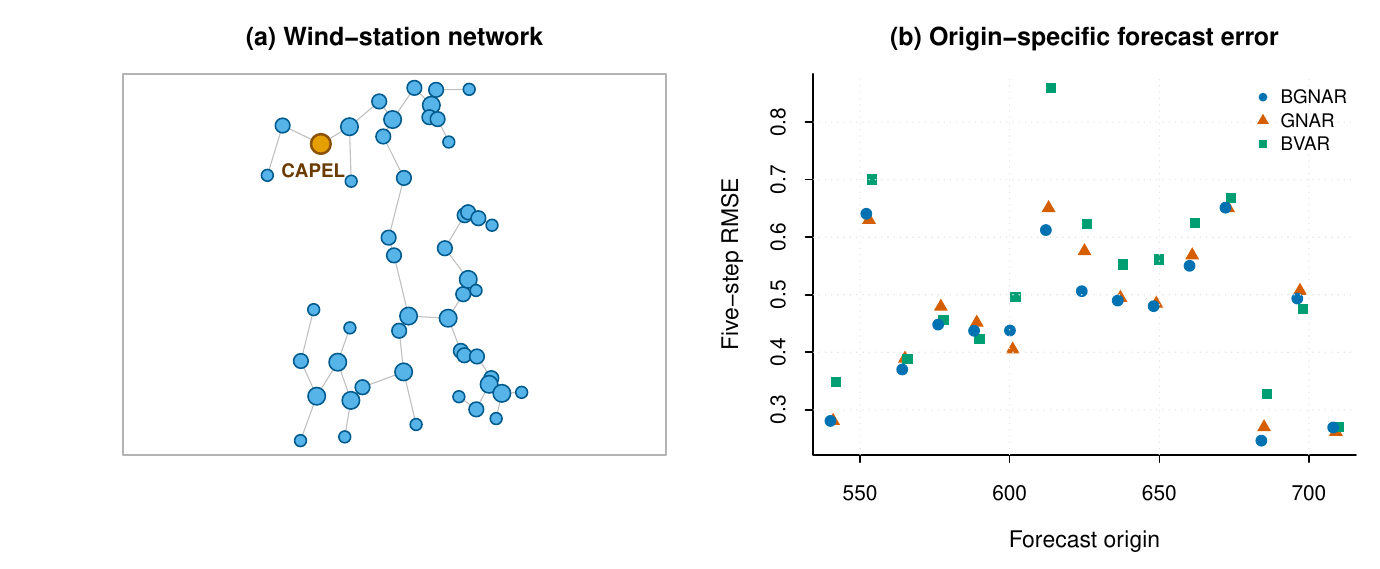}
\caption{Wind-speed network and forecasting performance.
(a) Network of the 50 selected weather stations.
(b) Five-step forecast RMSE for BGNAR, GNAR, and BVAR
at each of the 15 forecast origins.}
\label{fig:wind-analysis}
\end{figure}

We apply the proposed BGNAR model to the wind-speed time series and associated network provided by \cite{knight2020generalized}. The dataset consists of 721 observations from 102 weather stations in England and Wales, along with the network connectivity among stations. We select 50 connected stations using a breadth-first search starting from the CAPEL station. The resulting network contains 49 undirected edges, with node degrees ranging from one to three. The selected network and station locations are shown in Figure~\ref{fig:wind-analysis}(a). 

We conduct a rolling-window forecasting experiment using 240 observations for estimation and the subsequent five observations for forecast evaluation. We consider 15 forecast origins, starting at index 541 and ending at index 709, with consecutive origins spaced 12 observations apart.
As in Section~\ref{sec:simul}, BGNAR uses the fixed, prespecified temporal and neighborhood structure $(p,\bm s)=(5,[3,3,3,3,3])$, while BVAR uses a fixed temporal order of $p=5$. For GNAR, the temporal and neighborhood orders are selected by BIC within the maximum structure $(p,\bm s)=(5,[3,3,3,3,3])$. Across the 15 forecast origins, GNAR selects $(p,\bm s)=(2,[3,0])$ at six origins and $(p,\bm s)=(3,[3,0,0])$ at the remaining nine origins.

Figure~\ref{fig:wind-analysis}(b) shows the five-step forecast RMSE for each method across the 15 forecast origins, and Table~\ref{tab:wind-performance} summarizes the overall forecasting performance. BVAR has larger forecast errors at most origins. BGNAR achieves the smallest pooled RMSE, mean origin-specific RMSE, and MAE, although the differences from GNAR are modest. For predictive uncertainty, BGNAR achieves coverage of $0.926$, compared with $0.872$ for BVAR. Although the BGNAR intervals are wider, their coverage is closer to the nominal level of $0.95$, indicating more reliable predictive uncertainty quantification in this application.

\begin{table}[!t]
\centering
\caption{Forecasting performance for the wind-speed network application. Pooled RMSE is computed across all stations, forecast horizons, and origins. Mean (SD) denotes the mean and standard deviation of the 15 origin-specific RMSEs, and MAE denotes the mean absolute forecast error. Coverage and width refer to the empirical coverage and average width of 95\% posterior predictive intervals.}
\label{tab:wind-performance}
\footnotesize
\begin{tabular}{lrrrrr}
\toprule
Method & Pooled RMSE & Mean (SD) & MAE & Coverage & Width \\
\midrule
BGNAR & \textbf{0.477} & \textbf{0.461 (0.128)} & \textbf{0.330} & 0.926 & 1.650 \\
GNAR  & 0.490 & 0.473 (0.132) & 0.339 & NA & NA \\
BVAR  & 0.541 & 0.518 (0.160) & 0.368 & 0.872 & 1.428 \\
\bottomrule
\end{tabular}
\end{table}

\begin{figure}[!t]
\centering
\includegraphics[width=\textwidth]{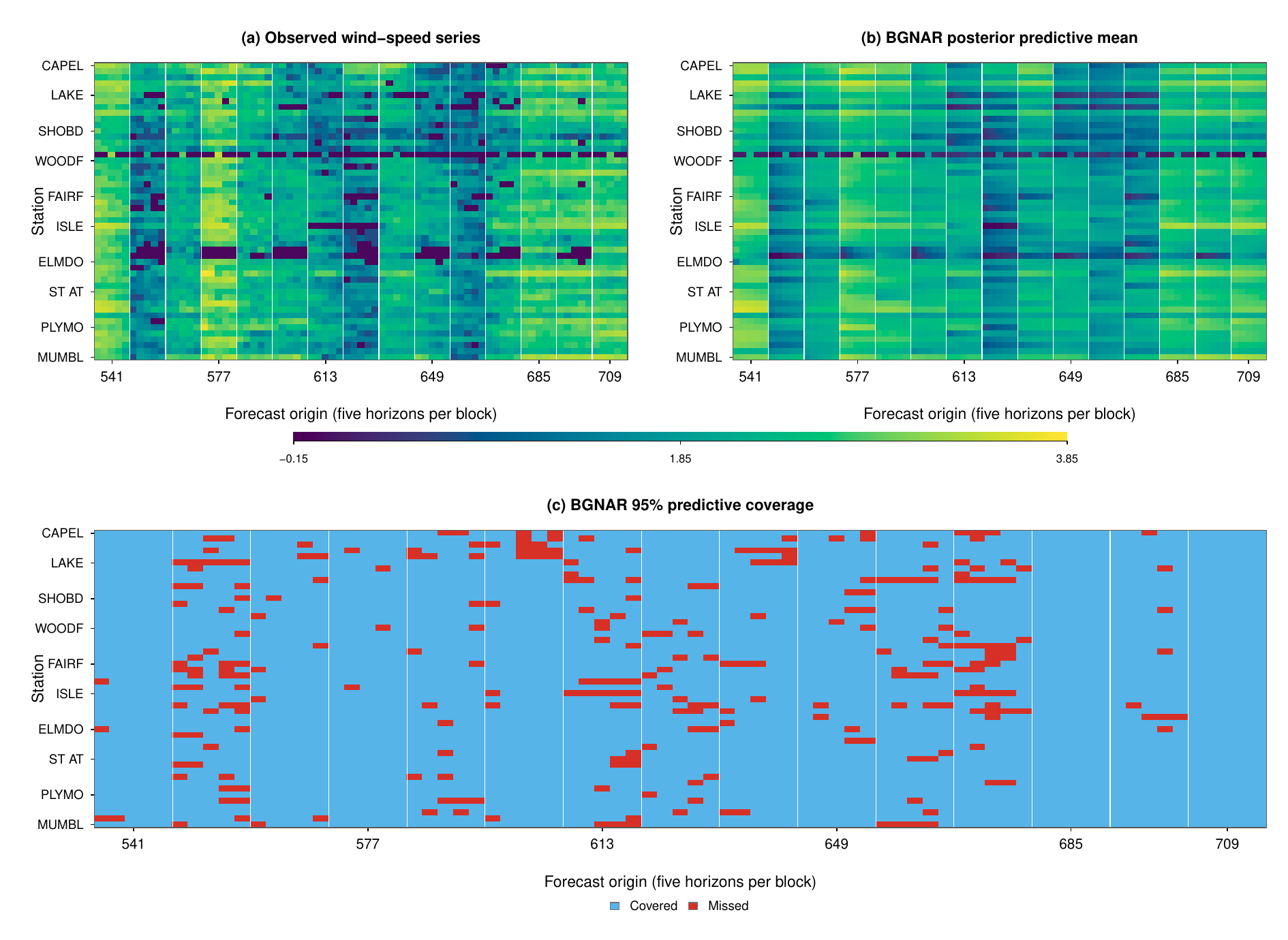}
\caption{Forecasting performance and 95\% posterior predictive interval coverage for BGNAR. (a) Observed wind-speed values for selected stations across forecast origins and horizons. (b) Corresponding posterior predictive means of BGNAR. (c) Indicators of whether the observed values fall within the 95\% posterior predictive intervals. }
\label{fig:wind-diagnostics}
\end{figure}

Figure~\ref{fig:wind-diagnostics} provides a station-level view of the BGNAR forecasts. Panels (a) and (b) show that the posterior predictive means generally follow the patterns of the observed wind-speed values across stations and forecast origins. Panel (c) shows that most observations fall within the corresponding 95\% posterior predictive intervals, consistent with the aggregate coverage reported in Table~\ref{tab:wind-performance}.

\begin{figure}[!t]
\centering
\includegraphics[width=\textwidth]{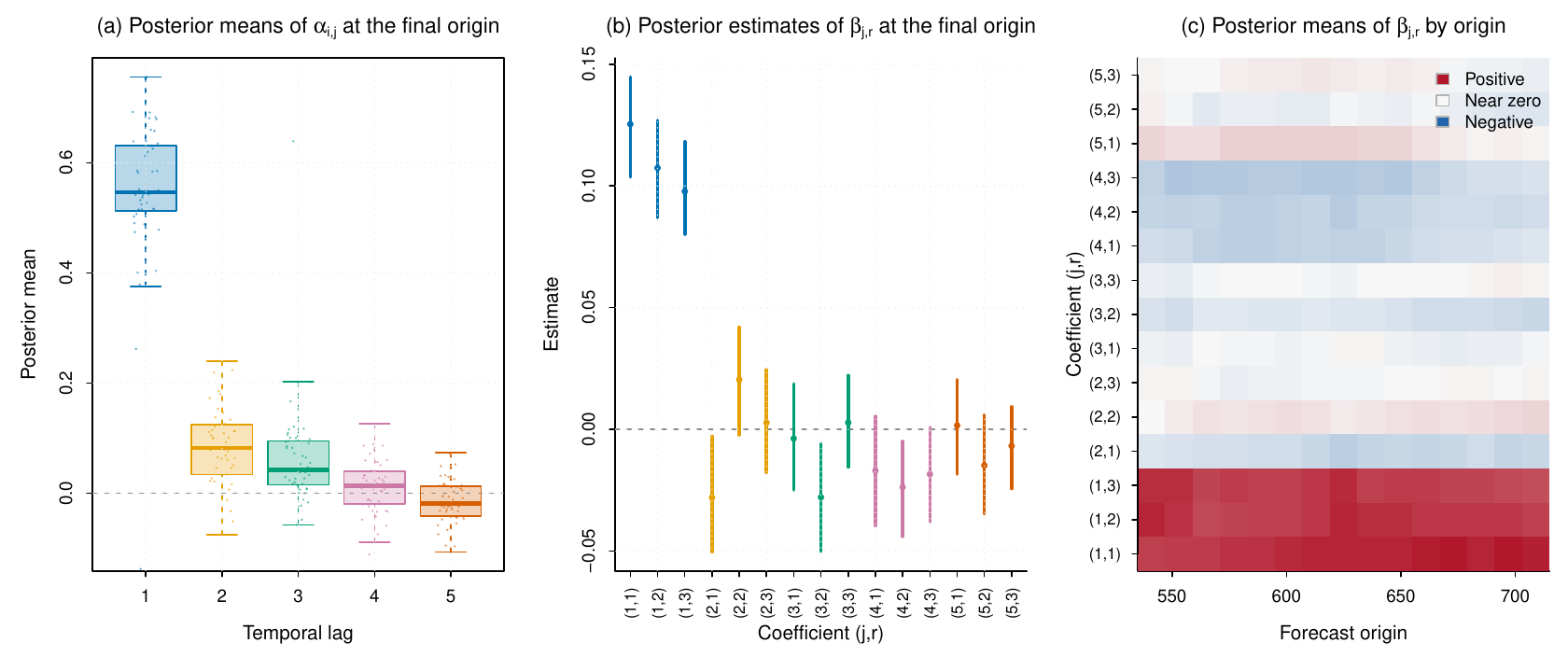}
    \caption{Posterior summaries of BGNAR coefficients in the wind-speed application.
(a) Across-station distributions of the posterior means of the station-specific own-lag coefficients $\alpha_{i,j}$ obtained from the final forecast origin. 
(b) Posterior means and 95\% credible intervals of $\beta_{j,r}$ at the final forecast origin. (c) Posterior means of $\beta_{j,r}$ across forecast origins. Each cell represents a coefficient $(j,r)$ at a given forecast origin.}
\label{fig:wind-parameters}
\end{figure}

Figure~\ref{fig:wind-parameters} summarizes the posterior estimates of the BGNAR coefficients. Panel (a) shows the across-station distributions of the posterior means of the own-lag coefficients at the final forecast origin, with each point representing one station's posterior mean at a given temporal lag. The own-lag effects are largest at the first lag and generally decrease at higher lags. Panel (b) shows the posterior means and 95\% credible intervals of the network-lag coefficients at the final forecast origin; the first-lag network effects are relatively large, while most coefficients at higher temporal lags are close to zero. Panel (c) shows the posterior means of $\beta_{j,r}$ across forecast origins, indicating that the first-lag network effects remain relatively large and positive across origins, whereas most higher-lag effects remain close to zero. Overall, these results are consistent with the intended shrinkage of higher-lag effects and provide interpretable posterior summaries of the network dynamics.

\section{Conclusion} \label{sec:conclusion}
We propose a Bayesian generalized network autoregressive (BGNAR) model for multivariate time series observed over a network. Building on the GNAR framework, we adapt priors commonly used in Bayesian vector autoregression (BVAR), including the Minnesota, sum-of-coefficients (SOC), and dummy-initial-observation (DIO) priors. The proposed Minnesota-type prior induces shrinkage across temporal lags and neighborhood orders, while the SOC and DIO priors incorporate persistence through dummy observations. The conditionally conjugate prior specification allows for posterior inference using a Gibbs sampler.
Simulation studies and the wind-speed network application demonstrate that BGNAR achieves forecasting performance competitive with GNAR and better than BVAR in the settings considered. BGNAR also provides stable predictive interval coverage across different sample sizes. In particular, it effectively shrinks unnecessary coefficients toward zero and maintains competitive forecasting performance even when the available time series is short. These results suggest that BGNAR provides a useful Bayesian approach to network time series forecasting, allowing uncertainty quantification and avoiding explicit lag-order selection through the use of a fixed over-specified structure. 

Several directions remain for future research. First, the Bayesian framework could be extended to jointly estimate the model coefficients and network weights, allowing the network structure to be learned from the data rather than treated as fixed. In particular, sparsity-inducing priors could be used to infer the presence or absence of network edges. Second, extending BGNAR to time-varying networks, where the network structure evolves over time, would broaden its applicability to dynamic network systems.

\section*{Acknowledgments}
\vspace{-3mm}
This work was supported by the National Research Foundation of Korea (NRF) grant funded by the Korea government (MSIT) (RS-2026-25471513), and by Basic Science Research Program and Global - Learning \& Academic research institution for Master's$\cdot$PhD students, and Postdocs(G-LAMP) Program through the NRF grant funded by the Ministry of Education (RS-2021-NR060140; No. RS-2026-25561016; No. RS-2025-25442252).

\appendix
\numberwithin{equation}{section}
\renewcommand{\thefigure}{\thesection.\arabic{figure}}
\renewcommand{\thetable}{\thesection.\arabic{table}}
\makeatletter
\@addtoreset{figure}{section}
\@addtoreset{table}{section}
\makeatother

\section{Explicit Form of $H_t$ in the GNAR Model}\label{sec:H_t}

Recall the GNAR model \eqref{eq:gnar_fix1} described in Section~\ref{subsec:gnar}.
\begin{equation} \label{eq:gnar_fix1_app}
    X_t = \mu + \sum_{j=1}^p \Phi_j X_{t-j} + u_t,
\end{equation}

To derive its linear regression representation, for each
temporal lag $j$, let
\[
\alpha_{\cdot,j}
=
(\alpha_{1,j},\ldots,\alpha_{N,j})^\top,
\qquad
\beta_j
=
(\beta_{j,1},\ldots,\beta_{j,s_j})^\top,
\]
and define
\[
\theta_j
=
\begin{pmatrix}
\alpha_{\cdot,j}\\
\beta_j
\end{pmatrix}
\in\mathbb R^{K_j},
\qquad K_j=N+s_j.
\]

For $i=1,\ldots,N$, let $J_i$ denote the $N\times N$ matrix with one in its
$(i,i)$th entry and zero elsewhere. Then, the coefficient matrix $\Phi_j$ satisfies
\[
\operatorname{vec}(\Phi_j) = \Gamma_j\theta_j,
\]
where
\[
\Gamma_j =
\left[
\operatorname{vec}(J_1), \ldots,
\operatorname{vec}(J_N),
\operatorname{vec}(W^{(1)}), \ldots,
\operatorname{vec}(W^{(s_j)}) \right]
\in\mathbb R^{N^2\times K_j}.
\]
Thus, at temporal lag $j$, an unrestricted VAR coefficient matrix contains $N^2$ unknown entries, whereas the corresponding individual-$\alpha$ GNAR coefficient matrix is determined by only $N+s_j$ parameters. Across all $p$ temporal lags, the individual-$\alpha$ GNAR model therefore contains $Np+\sum_{j=1}^p s_j$ dynamic coefficients, compared with $N^2p$ in an unrestricted VAR.

We stack the lag-specific parameters as
$\theta=(\theta_1^\top,\ldots,\theta_p^\top)^\top\in\mathbb R^{K_\theta}$,
where $K_\theta=\sum_{j=1}^p K_j$, and define
$\Gamma=\operatorname{blockdiag}(\Gamma_1,\ldots,\Gamma_p)$.
We further define
\[
\Psi=[\Phi_1,\ldots,\Phi_p]\in\mathbb R^{N\times Np},
\qquad
V_t=(X_{t-1}^\top,\ldots,X_{t-p}^\top)^\top\in\mathbb R^{Np}.
\]
Then, using the standard vectorization identity,
\[
\operatorname{vec}(\Psi)=\Gamma\theta,
\qquad
\Psi V_t=(V_t^\top\otimes I_N)\,\Gamma\theta.
\]
Defining $Z_t=V_t^\top\otimes I_N$, the GNAR model in (\ref{eq:gnar_fix1_app}) can therefore be written as
\begin{equation} \label{eq:gnar_z_gamma}
    X_t=\mu+Z_t\Gamma\theta+u_t.    
\end{equation}

With
\[
\gamma=(\mu^\top,\theta^\top)^\top
\in\mathbb R^{K_\gamma},
\qquad
K_\gamma=N+K_\theta,
\]
the design matrix is given by
\[
H_t=[I_N,Z_t\Gamma]
\in\mathbb R^{N\times K_\gamma}.
\]
Consequently, the GNAR model takes the linear
regression form
\begin{equation}
X_t=H_t\gamma+u_t,
\qquad t=p+1,\ldots,T.\notag
\end{equation}

\section{Design Matrix Construction for Dummy Observations}\label{sec:soc_dio}
Recall
\[
B=[\mu,\Phi_1,\ldots,\Phi_p]
\in\mathbb R^{N\times(1+Np)}
\]
and the matrix $\Gamma$ defined in Appendix~\ref{sec:H_t}.
Let
\[
G_\Gamma=\operatorname{blockdiag}(I_N,\Gamma).
\]
By the definition of $\Gamma$, we have
\[
\operatorname{vec}(B)=G_\Gamma\gamma.
\]
Using the standard vectorization identity,
\[
\operatorname{vec}(BY)
=(Y^\top\otimes I_N)\operatorname{vec}(B),
\]
the SOC design matrix is given by
\[
H_{\mathrm{soc}}
=
(Y_{\mathrm{soc}}^\top\otimes I_N)G_\Gamma
\in\mathbb R^{N^2\times K_\gamma},
\]
which satisfies
\[
H_{\mathrm{soc}}\gamma
=
\operatorname{vec}(BY_{\mathrm{soc}}).
\]

Similarly, the DIO design matrix is given by
\[
H_{\mathrm{dio}}
=
(Y_{\mathrm{dio}}^\top\otimes I_N)G_\Gamma
\in\mathbb R^{N\times K_\gamma},
\]
which satisfies
\[
H_{\mathrm{dio}}\gamma=BY_{\mathrm{dio}}.
\]

\section{Full Conditional Distributions and Gibbs Sampling Details}
\label{app:gibbs}
This section provides the explicit full conditional distributions used in the
Gibbs sampler described in Section~\ref{sec:posterior}. The notation follows Section~\ref{sec:model}.

\paragraph{Step 1. Sampling $\gamma$.}
Here, the dot denotes conditioning on the observed data and all remaining unknown quantities. The full conditional distribution of $\gamma$ is Gaussian:
\[
\gamma\mid\cdot
\sim N_{K_\gamma}(\bar\gamma,\bar V_\gamma),
\]
where $\bar V_\gamma=\bar Q_\gamma^{-1}$ and
\begin{align*}
\bar Q_\gamma
=
V_\gamma^{-1}
+H_{\mathrm{o}}^\top(I_{T_\ast}\otimes D_u^{-1})H_{\mathrm{o}}
+\tau_{\mathrm{soc}}^{-2}
(H_{\mathrm{soc}}^\ast)^\top H_{\mathrm{soc}}^\ast
+\tau_{\mathrm{dio}}^{-2}H_{\mathrm{dio}}^\top H_{\mathrm{dio}}.
\label{eq:gamma-precision}
\end{align*}
The corresponding full conditional mean is
\begin{align*}
\bar\gamma
=\bar V_\gamma\Big[
V_\gamma^{-1}m_\gamma(b)
+H_{\mathrm{o}}^\top(I_{T_\ast}\otimes D_u^{-1})x_{\mathrm{o}}
+\tau_{\mathrm{soc}}^{-2}
(H_{\mathrm{soc}}^\ast)^\top x_{\mathrm{soc}}^\ast
+\tau_{\mathrm{dio}}^{-2}H_{\mathrm{dio}}^\top x_{\mathrm{dio}}
\Big].
\end{align*}

\paragraph{Step 2. Sampling the error variances.}

Let
\[
e_t=X_t-H_t\gamma,
\qquad t=p+1,\ldots,T,
\]
and let $e_{i,t}$ denote its $i$th component. For each $i=1,\ldots,N$, the full conditional distribution is
\[
\sigma_i^2\mid\cdot
\sim
\mathrm{IG}\left(
c_\sigma+\frac{T_\ast}{2},\,
d_\sigma+\frac12\sum_{t=p+1}^{T}e_{i,t}^2
\right).
\]
Only the observed-data residuals enter this update, as the dummy-observation error variances are modeled separately.

\paragraph{Step 3. Sampling $b_j$.}

For each $j=1,\ldots,p$, the common own-lag component $b_j$ has the following Gaussian full conditional:
\[
b_j\mid\cdot
\sim
N\left(
\frac{\sum_{i=1}^{N}\alpha_{i,j}}{\kappa_\alpha+N}, \,
\lambda_\alpha^2j^{-a}
\frac{\kappa_\alpha}{\kappa_\alpha+N}
\right).
\]

\paragraph{Step 4. Sampling $\lambda_\alpha^2$.}

Define the own-lag quadratic form
\[
R_\alpha(\alpha,b,\kappa_\alpha;a)
=
\sum_{j=1}^{p}j^a
\left[
b_j^2+
\kappa_\alpha^{-1}
\sum_{i=1}^{N}(\alpha_{i,j}-b_j)^2
\right].
\]
Then, the full conditional distribution of
$\lambda_\alpha^2$ is
\[
\lambda_\alpha^2\mid\cdot
\sim
\mathrm{IG}\left(
c_\alpha+\frac{(N+1)p}{2},\,
d_\alpha+
\frac12R_\alpha(\alpha,b,\kappa_\alpha;a)
\right).
\]

\paragraph{Step 5. Sampling $\kappa_\alpha$.}

The full conditional distribution of $\kappa_\alpha$ is
\[
\kappa_\alpha\mid\cdot
\sim
\mathrm{IG}\left(
c_\kappa+\frac{Np}{2},\,
d_\kappa+
\frac{1}{2\lambda_\alpha^2}
\sum_{j=1}^{p}j^a
\sum_{i=1}^{N}(\alpha_{i,j}-b_j)^2
\right).
\]

\paragraph{Step 6. Sampling $\lambda_\beta^2$.}

Define the network-lag quadratic form
\[
R_\beta(\beta;a,d)
=
\sum_{j=1}^{p}\sum_{r=1}^{s_j}
j^ar^d\beta_{j,r}^2.
\]
Then, the full conditional distribution of
$\lambda_\beta^2$ is
\[
\lambda_\beta^2\mid\cdot
\sim
\mathrm{IG}\left(
c_\beta+\frac{\sum_{j=1}^{p}s_j}{2},\,
d_\beta+\frac12R_\beta(\beta;a,d)
\right).
\]

\paragraph{Step 7. Sampling $a$.}

The temporal-lag decay parameter $a$ is updated from its categorical full conditional distribution. For each $a_\ell\in\mathcal A$, compute
\begin{align*}
w_a(a_\ell)
={}&
\log\pi_a(a_\ell)
+\frac{a_\ell}{2}
\sum_{j=1}^p(N+1+s_j)\log j
\\
&-
\frac{1}{2\lambda_\alpha^2}
\sum_{j=1}^pj^{a_\ell}
\left[
b_j^2+
\kappa_\alpha^{-1}
\sum_{i=1}^N(\alpha_{i,j}-b_j)^2
\right]
\\
&-
\frac{1}{2\lambda_\beta^2}
\sum_{j=1}^p\sum_{r=1}^{s_j}
j^{a_\ell}r^d\beta_{j,r}^2.
\end{align*}
The corresponding posterior probabilities are
\[
P(a=a_\ell\mid\cdot)
=
\frac{\exp\{w_a(a_\ell)\}}
{\sum_{a'\in\mathcal A}\exp\{w_a(a')\}}.
\]
We sample $a$ from this categorical distribution.

\paragraph{Step 8. Sampling $d$.}

Similarly, the neighborhood-order decay parameter $d$ is updated from its categorical full conditional distribution. For each $d_m\in\mathcal D$, compute
\begin{align*}
w_d(d_m)
=
\log\pi_d(d_m)
+\frac{d_m}{2}
\sum_{j=1}^p\sum_{r=1}^{s_j}\log r
-
\frac{1}{2\lambda_\beta^2}
\sum_{j=1}^p\sum_{r=1}^{s_j}
j^ar^{d_m}\beta_{j,r}^2.
\end{align*}
The corresponding posterior probabilities are
\[
P(d=d_m\mid\cdot)
=
\frac{\exp\{w_d(d_m)\}}
{\sum_{d'\in\mathcal D}\exp\{w_d(d')\}}.
\]
We sample $d$ from this categorical distribution.

\paragraph{Step 9. Sampling the dummy tightness parameters.}

Define the residual vectors for the SOC and DIO dummy-observation blocks as
\[
r_{\mathrm{soc}}^\ast
=
x_{\mathrm{soc}}^\ast-H_{\mathrm{soc}}^\ast\gamma,
\qquad
r_{\mathrm{dio}}
=
x_{\mathrm{dio}}-H_{\mathrm{dio}}\gamma.
\]
The full conditional distribution of $\tau_{\mathrm{soc}}^2$ is
\[
\tau_{\mathrm{soc}}^2\mid\cdot
\sim
\mathrm{IG}\left(
c_{\mathrm{soc}}+\frac q2,\,
d_{\mathrm{soc}}
+\frac12(r_{\mathrm{soc}}^\ast)^\top r_{\mathrm{soc}}^\ast
\right).
\]
Similarly, the full conditional distribution of
$\tau_{\mathrm{dio}}^2$ is
\[
\tau_{\mathrm{dio}}^2\mid\cdot
\sim
\mathrm{IG}\left(
c_{\mathrm{dio}}+\frac N2,\,
d_{\mathrm{dio}}
+\frac12r_{\mathrm{dio}}^\top r_{\mathrm{dio}}
\right).
\]

\section{Hyperparameter Choices}\label{sec:hyperparameter}
We specify the hyperparameters governing the prior variances and dummy-observation tightness in BGNAR. Let $s_X$ denote the pooled standard deviation of the training observations across nodes and time points, which is used to calibrate scale-dependent prior hyperparameters.
For the intercept vector, we set
\[
\mu \sim N_N(0,c_\mu I_N),
\qquad c_\mu=10^7s_X^2.
\]
For the error variances and the node-specific variation
parameter, we use
\[
\sigma_i^2\sim \mathrm{IG}(2,s_X^2),
\qquad
\kappa_\alpha\sim \mathrm{IG}(3,2).
\]

For the two overall variance scales of the own-lag and network-lag coefficients, we assign
\[
\lambda_\alpha^2\sim \mathrm{IG}(3,0.16),
\qquad
\lambda_\beta^2\sim \mathrm{IG}(3,0.16).
\]
Each variance scale has a prior mode of $0.16/(3+1)=0.04$, whose square root is $0.2$. This choice is motivated by the default prior mode of $0.2$ for the Minnesota tightness parameter in the \texttt{BVAR} package of \cite{kuschnig2021bvar}.

For each dummy-observation block $k\in\{\mathrm{soc},\mathrm{dio}\}$, let $m_k$ denote its dimension, with $m_{\mathrm{soc}}=q$ and $m_{\mathrm{dio}}=N$. We assign
\[
\tau_k^2\sim \mathrm{IG}\left(
1+\frac{m_k}{2}\left(\frac{1}{b_\tau}-1\right),
\frac{a_\tau s_X^2m_k}{2b_\tau}
\right),
\]
where $a_\tau,b_\tau>0$ are chosen so that the inverse-gamma shape parameter is positive. Let $S_k$ denote the residual sum of squares of the corresponding dummy-observation block. The resulting full conditional mean is
\[
E(\tau_k^2\mid\cdot)
=
a_\tau s_X^2+b_\tau\frac{S_k}{m_k}.
\]
Here, $a_\tau s_X^2$ determines the baseline level of the conditional mean, while $b_\tau$ controls the contribution of the dummy residual sum of squares. We set $a_\tau=2.5$ and $b_\tau=1$, yielding
\[
\tau_{\mathrm{soc}}^2
\sim \mathrm{IG}\left(1,\frac{5qs_X^2}{4}\right),
\qquad
\tau_{\mathrm{dio}}^2
\sim \mathrm{IG}\left(1,\frac{5Ns_X^2}{4}\right).
\]

\section{Additional Simulation Results}\label{sec:additional_simul}
\begin{table}[!t]
\centering
\caption{Empirical coverage and average width of 95\% credible intervals for BGNAR coefficients. Results are reported separately for active (nonzero) and inactive (zero) coefficients, as well as for all coefficients.}
\label{tab:parameter-intervals}
\footnotesize
\setlength{\tabcolsep}{3.6pt}
\begin{tabular}{ccrrrrrr}
\toprule
$T$ & Parameter & \multicolumn{3}{c}{Coverage} & \multicolumn{3}{c}{Interval width} \\
\cmidrule(lr){3-5}\cmidrule(lr){6-8}
 & & Active & Inactive & All & Active & Inactive & All \\
\midrule
20 & $\alpha$ & 0.974 & 0.999 & 0.992 & 0.527 & 0.222 & 0.295 \\
   & $\beta$  & 0.803 & 0.999 & 0.984 & 0.370 & 0.176 & 0.195 \\
\addlinespace
40 & $\alpha$ & 0.985 & 0.998 & 0.994 & 0.382 & 0.161 & 0.214 \\
   & $\beta$  & 0.868 & 0.998 & 0.987 & 0.251 & 0.134 & 0.146 \\
\addlinespace
80 & $\alpha$ & 0.983 & 0.998 & 0.994 & 0.279 & 0.120 & 0.159 \\
   & $\beta$  & 0.908 & 0.995 & 0.987 & 0.177 & 0.105 & 0.112 \\
\addlinespace
160 & $\alpha$ & 0.982 & 0.997 & 0.994 & 0.207 & 0.091 & 0.119 \\
   & $\beta$  & 0.927 & 0.993 & 0.987 & 0.126 & 0.082 & 0.087 \\
\bottomrule
\end{tabular}
\end{table}

Table~\ref{tab:parameter-intervals} summarizes the empirical coverage and average width of 95\% credible intervals for the BGNAR coefficients. Coverage for active $\alpha$ coefficients remains above $0.95$ across all sample sizes, while coverage for active $\beta$ coefficients increases from $0.803$ at $T=20$ to $0.927$ at $T=160$. For inactive coefficients, coverage remains above $0.99$ for both $\alpha$ and $\beta$, indicating some overcoverage for coefficients that are truly zero. The interval widths decrease steadily as $T$ increases, reflecting lower posterior uncertainty with larger sample sizes. Overall, coverage for nonzero network effects falls below the nominal level in short samples but improves as the sample size increases, whereas the intervals for inactive coefficients tend to overcover.

\bibliographystyle{plainnat}
\bibliography{ref}

\end{document}